# Optics design of the Super Tau-Charm Facility collider rings

Y. Zou[1,*,†], L.H. Zhang[1,†], T. Liu[1], P.H. Yang[2], W.W. Li[2], T.L. He[2], D. Zhou[3], K. Ohmi[3], S.Y. Li[2], Z. Yu[1], Y.H. Mo[2], H.Z. Li[2], H. Zhou[1], J.J. Gao[1], Z.Y. Meng[1], Q. Luo[1,2], L. Wang[4], Y.J. Yuan[4], J.Y. Tang[1,2]

[1]*School of Nuclear Science and Technology, University of Science and Technology of China, No. 443 Huangshan Road, Hefei, 230027, China*

[2]*National Synchrotron Radiation Laboratory, University of Science and Technology of China, No. 42, South Cooperative Road, Hefei, 230029, China*

[3]*KEK, 1-1 Oho, Tsukuba 305-0801, Japan*

[4]*Institute of Modern Physics, Chinese Academy of Sciences, China*

[*]Corresponding author: zouye@ustc.edu.cn (Y. Zou)
[†]These authors (Y. Zou and L.H. Zhang) contributed equally to this work.

**ABSTRACT**

The Super Tau-Charm Facility (STCF), China's next-generation electron-positron collider, targets an unprecedented luminosity exceeding $5\times10^{34}$ cm$^{-2}$ s$^{-1}$ at a center-of-mass energy of 4 GeV. The implementation of a submillimeter vertical beta function at interaction point (< 1 mm) and crab-waist collision scheme in this low-energy regime introduces critical challenges through severe nonlinear effects that constrain dynamic aperture and degrade Touschek lifetime. To address these constraints, we propose a novel quasi-two-fold symmetric lattice design integrating several synergistic features: Linear optics optimization minimizing the H-invariant around the ring to maximize local momentum acceptance (LMA); Up to third-order of local chromaticity correction in the interaction region combined with second-order achromatic arc optics, enhancing off-momentum beam dynamics; Configured FODO arc structure with interleaved sextupole groups satisfying -I transformation, suppressing third-order geometric aberrations while optimizing Montague function distributions; Advanced final focus system integrating chromatic sextupoles, crab sextupoles, and strategically positioned octupoles to counteract final quadrupole fringe fields. Furthermore, we develop a multi-objective genetic algorithm using the in-house toolkit PAMKIT to simultaneously optimize 46 sextupole families, maximizing both dynamic aperture and momentum bandwidth. Optics performance is evaluated under error conditions with appropriate corrections, ensuring robust beam dynamics.

## 1. Introduction

The Super Tau-Charm Facility (STCF), hosted by the University of Science and Technology of China (USTC), is a new-generation double-ring electron-positron collider aimed at probing quark matter formation and fundamental interaction symmetries [1]. STCF targets tau lepton and charm quark pair production with symmetric $e^+e^-$ beam, operating at center-of-mass energies from 2 to 7 GeV. STCF adopts large Piwinski angle and crab waist collision scheme (CW), proposed by P. Raimondi in 2006 [2]. Tests at the DAΦNE Phi factory in 2010 [3] demonstrated that this scheme raises collision luminosity by one or two orders of magnitude without significant increase of beam intensity or decrease of bunch length. Recently, CW scheme becomes popular for circular $e^+e^-$ colliders [4]. Notably, SuperKEKB achieved a world-record luminosity of 5.1 × 10³⁴ cm⁻² s⁻¹ in 2024 using this scheme [5-7]. Newly designed circular $e^+e^-$ colliders, including the ultra-high-energy CEPC (IHEP, China) [8] and FCC-ee (CERN) [9-11], as well as the ultra-high-luminosity Tau/Charm Factory (INFN, Italy) [12] and Super Charm-Tau Factory (BINP, Russia) [13], all adopt this approach. Some previous studies about STCF collider rings can be found in Ref. [14-18].

STCF employs a 60 mrad crossing angle and the crab waist collision scheme, enabling a vertical beta function below 1 mm at the interaction point (IP). This design maintains vertical beam-beam tune shifts comparable to prior colliders while significantly enhancing luminosity at moderate beam currents, mitigating betatron resonances induced by beam-beam interactions [19-21]. The crab waist scheme requires crab sextupole magnets with precise phase advances relative to the IP, thereby introducing strong X-Y coupling to the beam in the interaction region (IR). Combined with the nonlinearities arising from strong focusing in the IR, this reduces dynamic and momentum apertures, complicating beam injection and yielding a Touschek lifetime below 300 s, which is much shorter than typical for ring-based accelerators, at high bunch intensities and low emittance. This necessitates a high-repetition-frequency injector, posing design challenges. STCF adopts a hybrid injection scheme, initially using off-axis injection with provisions for upgrading to swap-out injection. Table I summarizes the overall parameters for STCF collider rings.

Table 1: The overall design parameters for the STCF collider rings

| Parameter | Value |
| --- | --- |
| Beam energy (GeV) | 1~3.5 |
| Luminosity (cm⁻²s⁻¹) | ⩾5×10³⁴(@2 GeV) |
| $\beta_y^*$ (mm) | ⩽1(@2 GeV) |
| Touschek lifetime (s) | >200 |
| Circumference (m) | 800~1000 |

This paper presents an overview of the latest optics design of STCF and is organized as follows: Section II outlines the layout of collider rings. Section III discusses the interaction region design. Section IV introduces the arc section design. Section V details the straight section design. Section VI discusses beam dynamics and nonlinear optimization. Section VII evaluates lattice performance and its sensitivity to errors.

## 2. LAYOUT

The STCF collider rings adopt a quasi-two-fold symmetric double-ring configuration. Each ring, housing electrons or positrons in a common horizontal plane, has a circumference of 860.321 m and intersects at the IP and a diametrically opposite point. The arc sections feature slightly asymmetric drift lengths, forming alternately half inner and outer rings separated by about 2 m.

Each ring comprises one interaction region, four 60° major arc sections, two 30° minor arc sections, one intersection region, and multiple straight sections for injection, extraction, damping wigglers, RF cavities, and beam collimation. Key lattice design features include:

- A symmetric layout is to preserve the option of adding a second IP and is beneficial to beam dynamics.
- A special arc section of 60° bending angle between crab sextupoles is much helpful for local chromaticity correction, and a very weak bending magnet close to IP is to minimize

- synchrotron radiation to the vacuum chamber around the IP.
- A relatively small bending angle of 60° for both the e+e- injection beamline is helpful to control the emittance growth, as compared to BEPCII (90°) and SuperKEKB (~90° and >180°).
- A modest length of 60° for the main arc sections is found helpful in optimizing local chromaticity correction and overall dynamic/momentum apertures.
- Reserved straight sections around the IP are for installing spin rotators in the potential upgrading for polarized electron beam.

Figure 1 illustrates the lattice layout and optics functions.

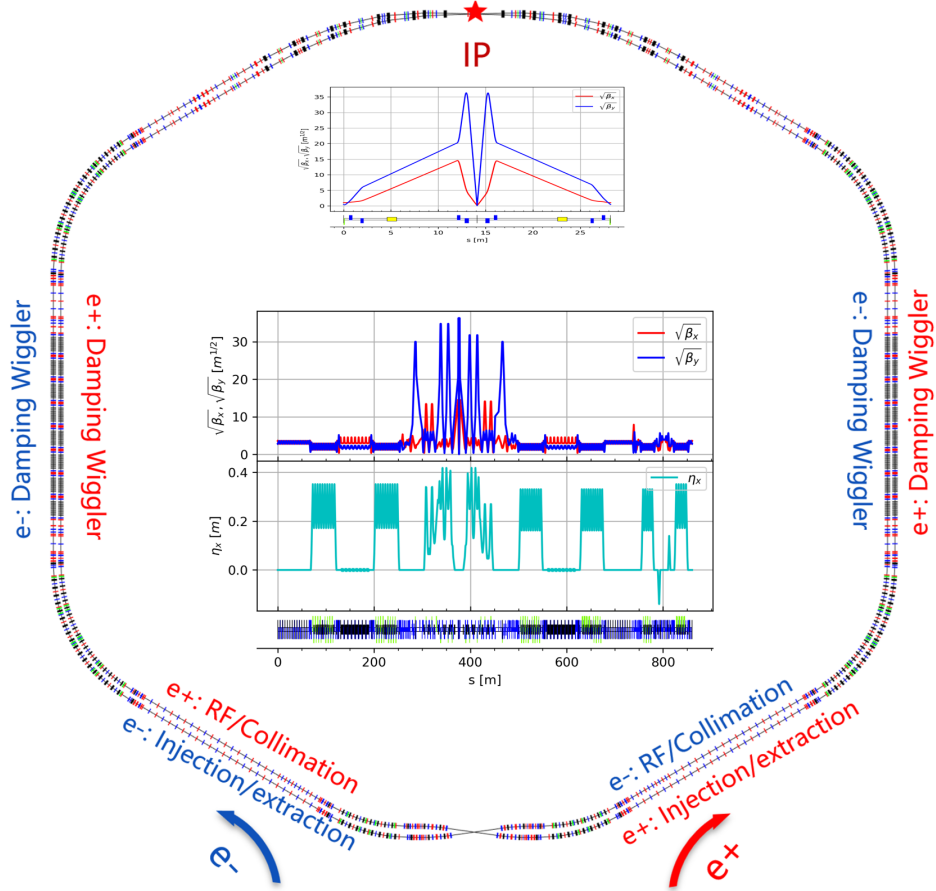

Figure 1: The layout and optics function of the STCF collider rings.

## 3. INTERACTION REGION

The STCF IR lattice is designed to squeeze the beta function at the IP to maximize luminosity while ensuring sufficient dynamic and momentum apertures for an adequate Touschek lifetime. This requires optimized linear optics and nonlinear corrections, coordinating phase advances for nonlinear cancellation and minimizing sextupole field effects.

Following design principles of modern $e^+e^-$ colliders [4,5], the IR employs a modular layout (Fig. 2) comprising: (1) a final focus telescope (FFT) with superconducting and normal-conducting quadrupole doublets to achieve a vertical beta function of <1 mm and form an IP image point for higher-order chromaticity correction; (2) vertical and horizontal chromaticity correction sections (CCY/CCX) with -I transformation to cancel sextupole nonlinearities, using two π/2-phase-advance FODO cells and symmetric dipoles; (3) matching sections (MCY, YMX) for beta function and dispersion control, adjusting phase advances between main sextupole pairs S1Y/S1X and final focus quadrupoles to correct second-order chromaticity; (4) a dispersion suppressor (XMC) to nullify dispersion at the crab sextupole section (CS); and (5) a matching section (MS) linking the IR to the straight section. The CS satisfies phase advance constraints ($\mu_x = 6\pi$, $\mu_y = 5.5\pi$, $\alpha_x = 0$, $\alpha_y = 0$) and ensures $\beta_y \gg \beta_x$ to reduce crab sextupole strength, mitigating nonlinear effects.

The IR, spanning ~210 m with a 60° deflection, uses 1 m dipoles optimized for large dispersion at sextupole pairs (reducing sextupole strength) while maintaining a dispersion invariant $H_x < 0.02$ m to enhance local momentum acceptance. To achieve a 60 mrad crossing angle, inner ring dipoles bend 30 mrad less and outer ring dipoles 30 mrad more, maintaining a 1.5–2 m ring separation and asymmetric dispersion about the IP. Figure 2 illustrates the IR layout and optics functions.

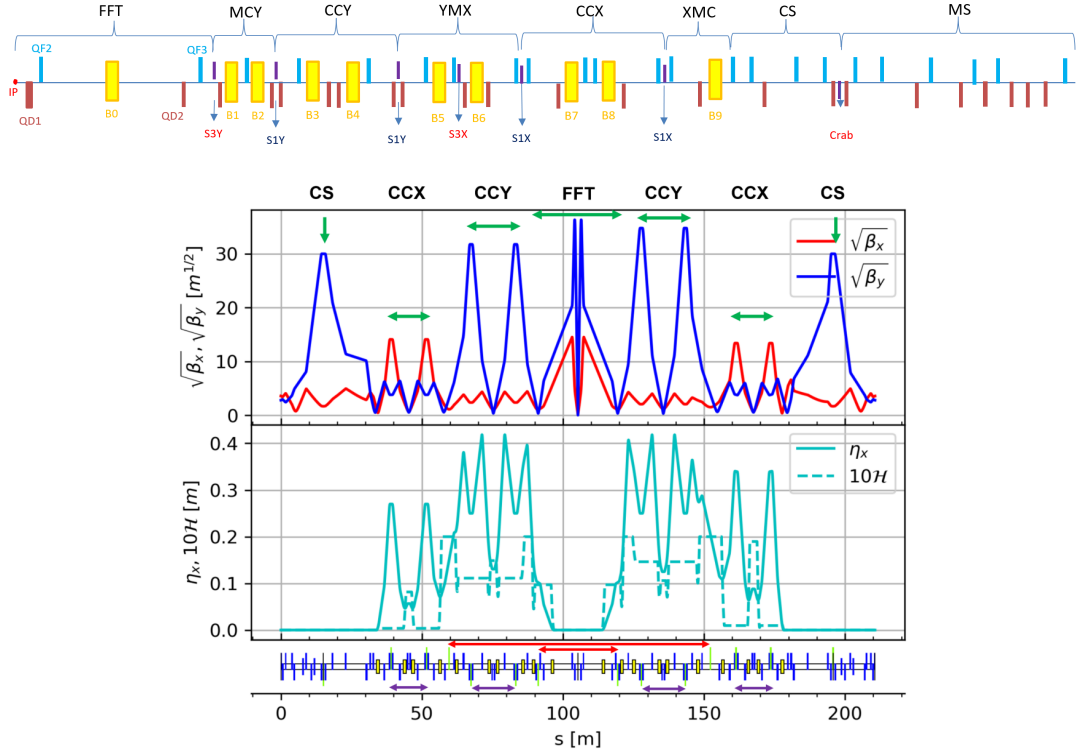

Figure 2: The layout (upper) and optics function (lower) of interaction region (The purple arrows indicate sextupole pairs to correct 1st order chromaticity; the red arrows indicate sextupoles to correct 3rd order chromaticity)

## 4. ARC SECTIONS

The arc sections are designed to achieve low emittance and a large momentum compaction factor ($\alpha_c$) to mitigate beam instabilities and satisfy the coherent X-Z instability requirement, ensuring the horizontal beam-beam parameter is significantly less than the synchrotron tune. Each arc employs 90°/90° FODO cells, comprising one focusing quadrupole, one defocusing quadrupole, and two bending magnets.

Figure 3 illustrates the scaling of emittance ($\varepsilon_0$), momentum compaction factor ($\alpha_c$), and phase advances ($\mu_x, \mu_y$) with drift length, quadrupole strength, bending length, and angle. Smaller bending angles reduce emittance but also decrease $\alpha_c$, requiring a balance with dispersion, chromaticity, $H$ invariant, and beta functions. The optimized FODO cell, with a length of 4.7 m and a 6° bending angle, yields a maximum beta function of 7.87 m, dispersion of 0.33 m, and H invariant of 0.0186 m (Fig. 4), maximizing momentum acceptance.

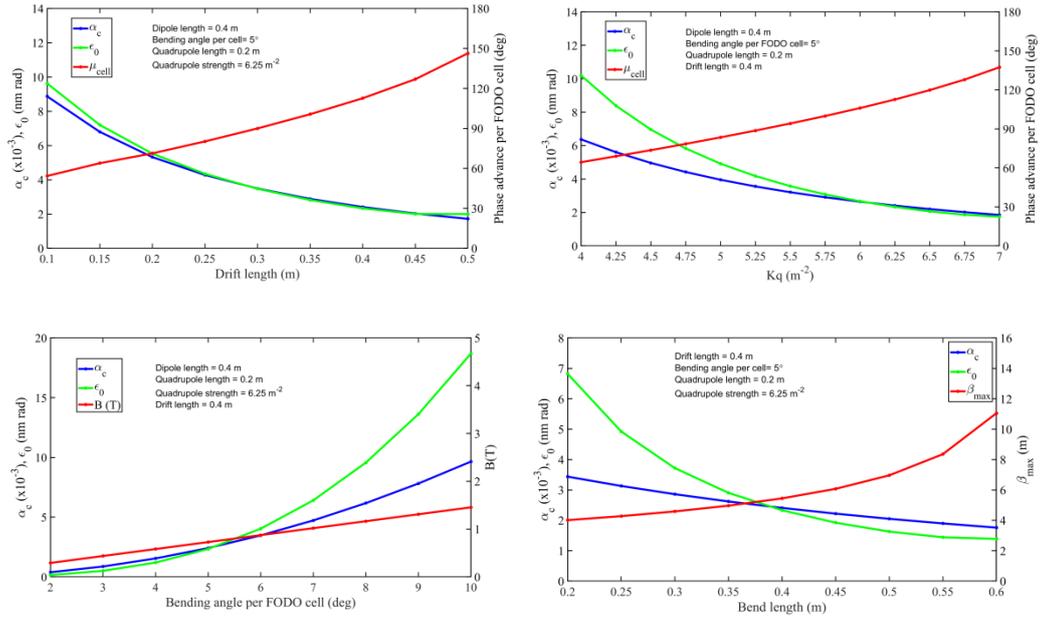

Figure 3: Scaling of emittance ($\varepsilon_0$), momentum compaction factor ($\alpha_c$), and phase advances ($\mu_x$, $\mu_y$) with drift length, quadrupole strength, bending length and angle for an FODO lattice

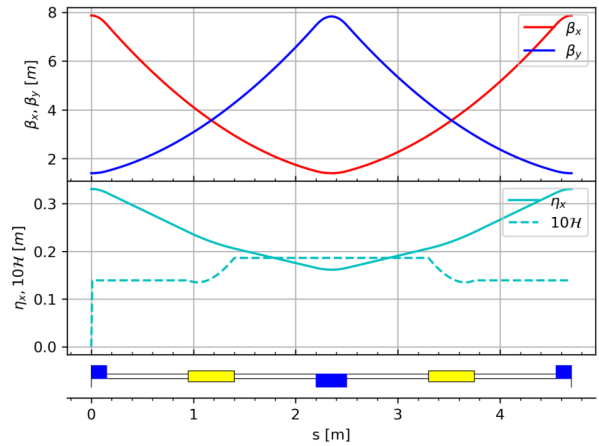

Figure 4: Layout and optics function of the FODO cell for the arc

The long and short arc sections comprise 9 and 4 FODO cells, respectively. Sextupoles, positioned immediately following quadrupoles (SF after focusing quadrupoles, SD after defocusing quadrupoles), facilitate chromaticity correction and nonlinear optimization. Sextupole pairs, separated by a 180° phase advance, satisfy -I transformations to cancel first-order geometric resonance terms. The long arc section contains 8 sextupole groups (4 SF, 4 SD), while the short arc section has 4 groups (2 SF, 2 SD), as shown in Fig. 5. This configuration, inspired by the second-order achromat concept [22,23], allows independent adjustment of sextupole field strengths, enhancing flexibility for global nonlinear optimization.

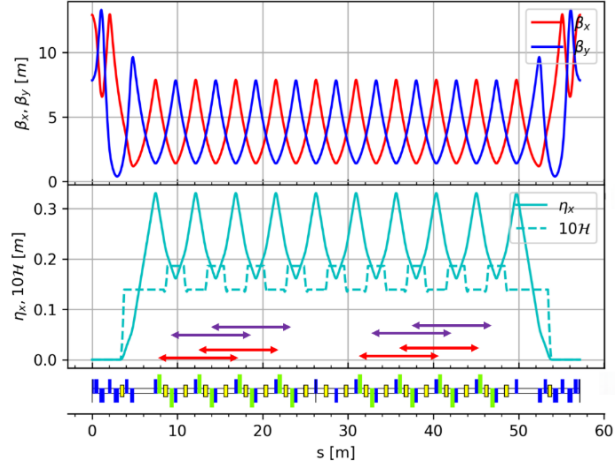

Figure 5: Optics function of long arc section (Red arrows indicate SFs in pairs and purple arrows indicate SDs in pairs)

## 5. STRAIGHT SECTIONS

STCF collider rings incorporate several straight sections, including general straight sections and specialized segments for injection, damping wigglers, each tailored to specific functions. This section presents the optics design of the general straight sections, alongside the injection and damping wiggler sections, ensuring compatibility with the ring's overall lattice and performance objectives.

*General straight section*

The general straight section of the STCF collider rings employ a FODO lattice with eight 5 m cells, each featuring a 30° phase advance and two 2.2 m drift spaces. Figure 6 illustrates the optical functions of 4 FODO cells in the general straight section.

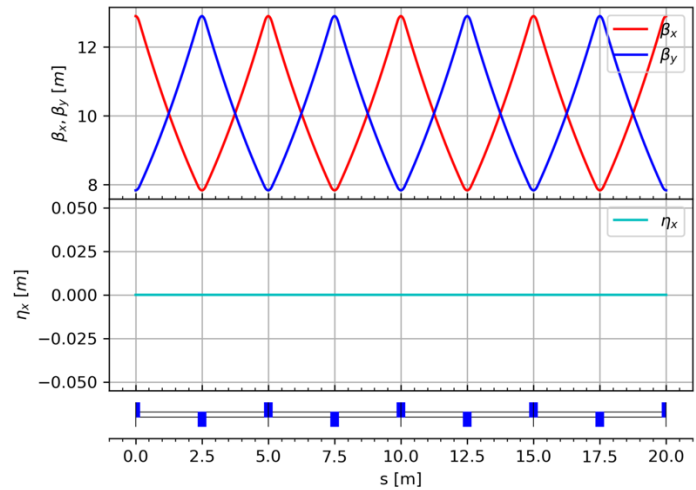

Figure 6: Optical functions of 4 FODO cells in the general straight section

*Beam injection and extraction section*

The primary goal of the injection system design is to achieve high-efficiency injection of electron and positron beams from the injector, ensuring sustained high integrated luminosity while minimizing perturbations to the circulating beam and mitigating the impact of injection-related beam losses on detector background.

To balance the physical requirements of the injector and overall project cost, the STCF collider rings adopt a hybrid injection scheme compatible with both off-axis and swap-out injection methods. In the initial phase, an off-axis injection scheme is implemented, utilizing septum and bump magnets. In this approach, the injected beam is transversely offset from the circulating beam, with the latter's closed orbit shifted toward the septum magnet via bump magnets. This mature scheme requires less stringent pulse timing for injection components but necessitates a sufficiently large dynamic aperture to accommodate the beam offset. In the subsequent phase, the system will transition to a swap-out injection scheme, where pulsed kicker magnets replace circulating beam bunches with new injected bunches. This method reduces the demand on horizontal dynamic aperture but requires kicker magnet pulse widths to be less than twice the minimum bunch spacing. Additionally, injecting high-charge positron bunches imposes significant demands on the repetition frequency of the upstream injector.

To facilitate a seamless upgrade from off-axis to swap-out injection, the design incorporates a unified quadrupole magnet layout, requiring only the replacement of kicker components. Comprehensive tracking simulations and optimized hardware parameter selections have been conducted to validate the injection process and ensure compatibility with both schemes. Figure 7 shows the optics design of the off-axis and swap-out injection schemes.

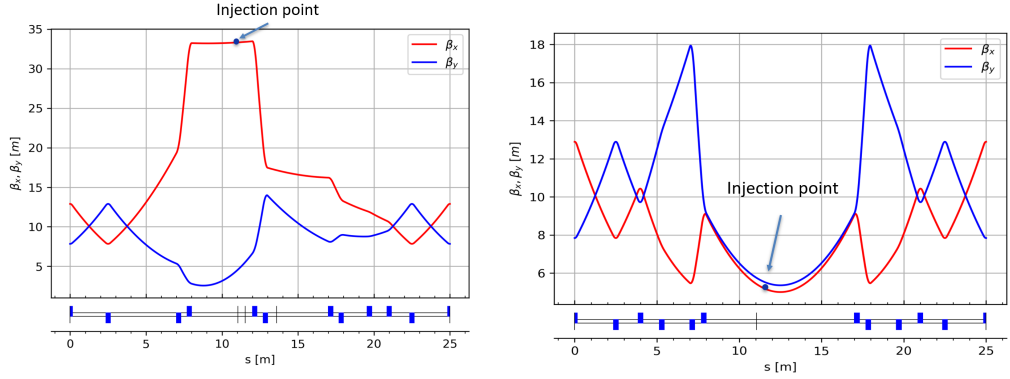

Figure 7: The optics function of injection schemes (Left: off-axis injection; Right: swap-out injection)

*Damping wiggler sections*

Each of the STCF collider rings incorporates two damping wiggler sections to reduce damping time and adjust emittance across the energy range of 1.0–3.5 GeV. Each section comprises eight 4.8 m damping wiggler magnets within a triplet lattice configuration, optimized to ensure smooth beta function variation and accommodate long drift spaces. Quadrupole magnets at section ends correct optical perturbations induced by the wigglers, matching Twiss parameters to adjacent straight sections. Due to the absence of realistic s-dependent magnetic field models in standard lattice design tools (e.g., MADX [24], SAD [25]), wigglers are approximated as sequences of bending magnets and drift spaces to replicate radiation damping effects. Figure 8 illustrates the optics functions of the damping wiggler section.

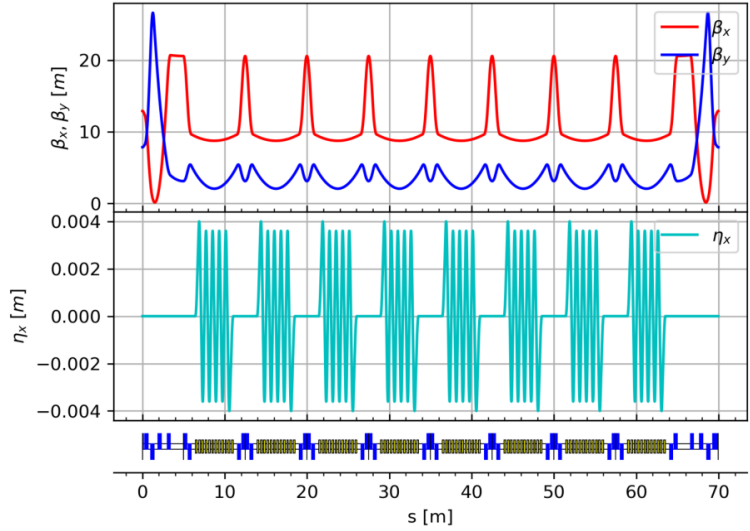

Figure 8: The optics function of damping wiggler section

## 6. LOCAL CHROMATICITY CORRECTION AND NONLINEAR OPTIMIZATION

Chromaticity correction and nonlinear optimization are critical challenges in collider rings design. Local chromaticity correction up to the third order is implemented at the interaction region (IR) as follows: (1) First-order chromaticity is corrected using non-interleaved horizontal and vertical sextupole pairs (S1X and S1Y, Fig. 2) with a -I transformation; (2) Second-order chromaticity is tuned by adjusting phase advances between sextupoles and final focus (FF) quadrupoles; and (3) Third-order chromaticity is corrected using sextupoles (S3X and S3Y, Fig. 2) at IP image points with small β-functions. Chromaticity in non-IR regions is corrected using sextupole pairs in the arc sections. Crab sextupoles are placed in a nearly achromatic region to reduce their impact on the momentum aperture.

The strong final focusing quadrupoles (QD1 and QF2) cause large Montague chromatic functions ($W_y$, $W_x$), which describe optics perturbations for off-momentum particles [26,27]. The horizontal and vertical W functions originate from QF2 and QD1, respectively. By setting the phase advances between QD1 and S1Y, and between QF2 and S1X, to (π, 3π), $W_y$ and $W_x$ are significantly reduced. Optimizing the strength of S1X and S1Y further minimizes the W function at the crab sextupoles. Additionally, phase advances between the first defocusing sextupole (SD) in the arc and QD1, and between the first focusing sextupole (SF) in the arc and QF2, are set to integer multiples of π. This configuration renders the non-IR sections "transparent" to the W function, maintaining a low W value in these regions. Figure 9 illustrates the Montague chromatic function across the entire ring, with the one-turn W function reduced to approximately 10 and $\delta W_{x,y}$ below 0.01 (where $\delta$ is the energy spread).

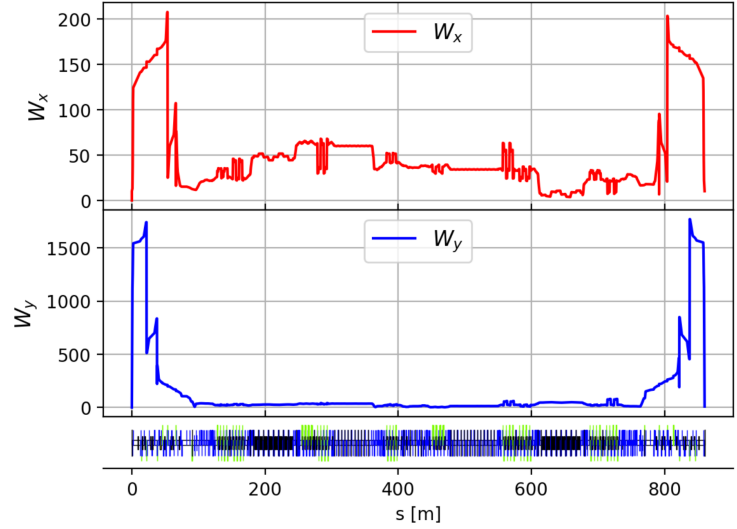

Figure 9: The Montague chromatic function along the whole ring

Multi-objective optimization via in-house developed PAMKIT code [28] optimizes dynamic and momentum apertures using 46 sextupole groups (6 in the IR, 40 in the arcs) as tuning knobs. Optimization objectives are momentum aperture, amplitude driving terms, and dynamic aperture. During optimization, crab sextupoles are always activated. The strength of sextupoles in the IR slightly vary with respect to the set values for local chromaticity correction of IR, while the strength of sextupoles in the arcs can significantly vary to achieve the optimization objectives.

After optimization, SAD is used for the precise evaluation of the collider rings' dynamic aperture by 6D multipole particle tracking, which involves synchrotron radiation in all magnets, high-order kinematic effects, finite-length effects of sextupoles, and Crab sextupoles. Figure 10 illustrates the dynamic apertures with crab sextupoles enabled and disabled, showing a slight reduction in on- and off-momentum apertures with crab sextupoles active, yet maintaining dynamic apertures $>30\sigma_x$ horizontally and $>120\sigma_y$ vertically, with $>15\sigma_x$ at $10\sigma_\delta$ momentum deviation. The momentum acceptance exceeds $\pm 1.4\%$ (seeing Fig. 11).

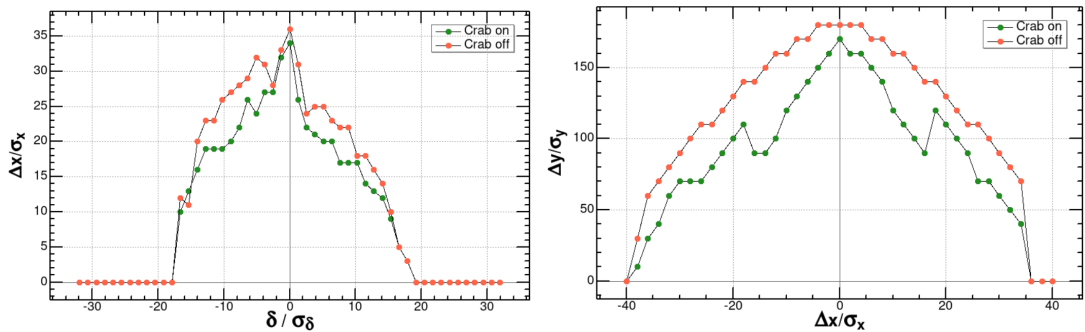

Figure 10: Off-momentum (left) and on-momentum (right) dynamic apertures with crab sextupoles on and off.

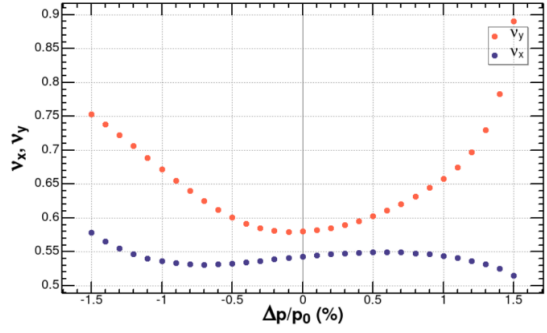

Figure 11: Tune (fractional part) versus momentum deviation δ.

Fringe fields, particularly those of final focus (FF) quadrupoles, significantly influence nonlinear beam dynamics in the STCF collider rings. The Hamiltonian for the quadrupole fringe field is given by [29]:

$$H = -\frac{k_1'(s)x^2 y p_y}{2} + \frac{k_1''(y^4 - 6x^2 y^2)}{24},  \quad (1)$$

where $k_1'(s) = B_1'(s)/B\rho$, $k_1''(s) = B_1''(s)/B\rho$. To counteract these nonlinear effects, octupole coils are installed external to the FF quadrupoles. Figure 12 illustrates the on- and off-momentum dynamic apertures with and without FF quadrupole fringe fields and octupole compensation, with crab sextupoles enabled. The results show a substantial reduction in dynamic aperture due to FF quadrupole fringe fields, partially mitigated by octupole coils. Ongoing optimization of octupole configurations aims to further enhance aperture recovery.

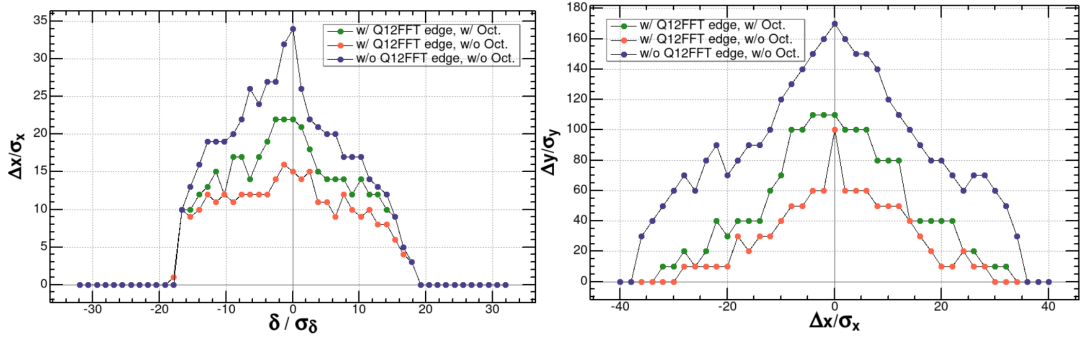

Figure 12: Dynamics apertures with and without FF quadrupole fringe fields octupole compensation (Left: off-momentum dynamic aperture; Right: on-momentum dynamic aperture)

Frequency map analysis (FMA) [30] is employed to evaluate the chaotic behavior of the beam dynamics. It can apply to Hamiltonian system or symplectic map. The tune spread is expressed as:

$$D = \log_{10}\sqrt{(v_{x2} - v_{x1})^2 + (v_{y2} - v_{y1})^2},  \quad (2)$$

where $(v_{x1}, v_{y1})$ are betatron tunes calculated from the first 1000 turns and $(v_{x2}, v_{y2})$ from the second 1000 turns of a particle tracking. Figure 13 illustrates the dynamic aperture and tune spread for the case with final focus (FF) quadrupole fringe fields and octupole compensation. The results indicate the absence of low-order resonance lines within the dynamic aperture, confirming stable beam dynamics.

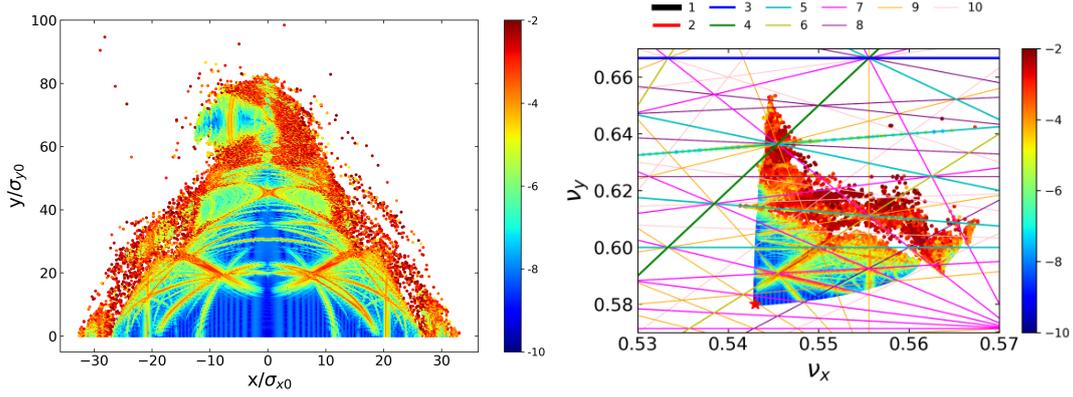

Figure 13: Dynamic aperture (left) and tune spread (right) for the case with FF quadrupole fringe field and octupole compensation. (The color bar represents the diffusion rate with blue color for regular motion and red for chaotic)

The local momentum acceptance (LMA) [31], denoted as $\delta_n(s)$ and $\delta_p(s)$, represents the maximum negative and positive fractional momentum deviations a particle can undergo at location s while remaining stable after a specified number of turns. LMA is a key factor in determining the Touschek lifetime. It can be expressed as [13],

$$\delta(s) = \frac{R}{\eta(s)+\sqrt{\mathcal{H}(s)\beta_x(s)}} \qquad (3)$$

where $R$ is the aperture limitation, $\eta(s)$ is the dispersion function, $\mathcal{H}(s)$ is the H-invariant, and $\beta_x(s)$ is the beta function. To increase LMA, one can enlarge the dynamic aperture or reduce the dispersion function, beta function, or H-invariant. However, the dynamic aperture is constrained by strong nonlinear effects, and the beta and dispersion functions must balance chromaticity correction and nonlinear optimization. Minimizing the H-invariant is an effective strategy to enhance LMA. During the linear optics design of the ring, we aim to minimize the H-invariant to maximize LMA. As shown in Figure 14, the H-invariant is kept below 0.02 m across the ring.

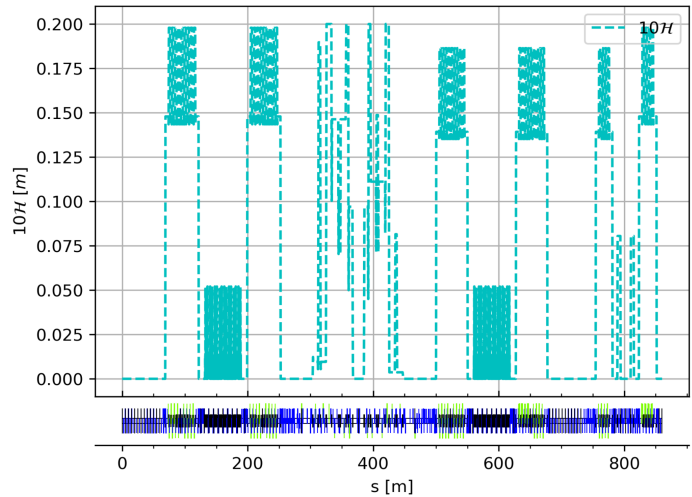

Figure 14: The H invariant function across the ring

The Elegant [32] code was employed to track LMA profiles across the STCF collider rings for cases with and without FF quadrupole fringe fields and octupole compensation. Figure 15 shows that FF quadrupole fringe fields significantly reduce LMA compared to the idealized fringe-field-free case, with the interaction region identified as the primary constraint on global momentum acceptance. Octupole compensation substantially improves LMA, achieving minimum values of approximately +0.7% and −0.6% for positive and negative momentum deviations, respectively. At a beam energy of 2 GeV and luminosity of $1\times10^{35}$ cm$^{-2}$s$^{-1}$, the Touschek lifetime reaches at least 300 s with the optimized lattice, accounting for fringe field effects and compensation, meeting STCF design requirements. Table 2 shows the main design parameters of the STCF collider rings.

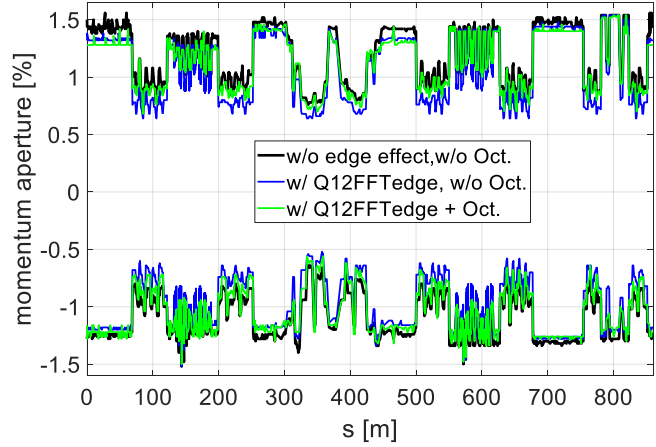

Figure 15: Local momentum acceptance w/ and w/o fringe fields effect and with/without octupoles correction.

Table 2: The main parameters of the STCF collider rings

| Parameter | Value | Value | Value | Value |
|---|---|---|---|---|
| Beam energy (GeV) | 2.0 | 1.0 | 1.5 | 3.5 |
| Circumference (m) | 860.321 | 860.321 | 860.321 | 860.321 |
| Beam current (A) | 2.0 | 1.1 | 1.7 | 2.0 |
| Crossing angle (mrad) | 60 | 60 | 60 | 60 |
| $\beta_x^*/\beta_y^*$ (mm) | 60/0.8 | 60/0.8 | 60/0.8 | 60/0.8 |
| Coupling, $\varepsilon_y/\varepsilon_x$ (%) | 1 | 1 | 1 | 0.5 |
| Hor./Ver. betatron tune | 30.543/34.58 | 30.55/34.57 | 30.55/34.57 | 30.55/34.57 |
| Hor. Emit. (SR/DW+IBS) (nm) | 8.79/4.63 | 2.2/5.42 | 4.94/3.82 | 26.9/26.91 |
| Momentum compact factor ($10^{-3}$) | 1.35 | 1.26 | 1.32 | 1.37 |
| Energy spread (DW+IBS) ($10^{-4}$) | 7.8 | 6.18 | 6.93 | 10.02 |
| Energy loss per turn (keV) | 543 | 106 | 267 | 1494 |
| SR power per beam (MW) | 1.086 | 0.117 | 0.453 | 2.988 |
| RF voltage (MV) | 2.5 | 0.75 | 1.2 | 6 |
| Synchrotron tune | 0.0194 | 0.0146 | 0.0154 | 0.0228 |
| Energy acceptance (%) | 1.68 | 1.44 | 1.35 | 1.88 |
| Bunch length (DW+IBS) (mm) | 7.6 | 7.05 | 8.18 | 8.26 |
| Beam-beam parameter, $\xi_x/\xi_y$ | 0.005/0.095 | 0.005/0.023 | 0.004/0.033 | 0.003/0.032 |
| Luminosity (cm$^{-2}$s$^{-1}$) | 9.4×10$^{34}$ | 6.19×10$^{33}$ | 2.09×10$^{34}$ | 4.48×10$^{34}$ |

## 7. SENSITIVITY TO ERRORS

The lattice design of the STCF collider rings have been analyzed to evaluate its robustness against alignment errors, which are critical for maintaining beam stability and achieving high luminosity. Figure 16 presents the orbit survival ratio as a function of misalignment in the FFT doublets, which are crucial for focusing the beam at the interaction point. Figure 17 illustrates the vertical beta-beating induced by misalignments of the crab sextupoles in the IR. Simulation results reveal that the orbit survival ratio is highly sensitive to FFT doublet offsets, with even 50 μm misalignments causing significant reductions in beam stability. Similarly, the vertical beta-beating exhibits pronounced sensitivity to crab sextupole offsets in the IR, leading to substantial distortions in the vertical beam optics. These findings highlight the critical need for stringent alignment tolerances for both the FFT doublets and crab sextupoles to ensure optimal performance of the STCF collider. Table 3 summarizes the reference alignment error tolerances for key components of the STCF collider rings, providing a benchmark for mechanical design and installation precision. To mitigate the impact of these sensitivities, ongoing studies are exploring advanced correction techniques, including orbit feedback systems and dynamic alignment adjustments. Additionally, optimization of the lattice design is being pursued to reduce the sensitivity to misalignments while balancing other performance constraints, such as chromaticity correction and nonlinear dynamics. These efforts aim to enhance the operational reliability and efficiency of the STCF collider, ensuring it meets its scientific objectives.

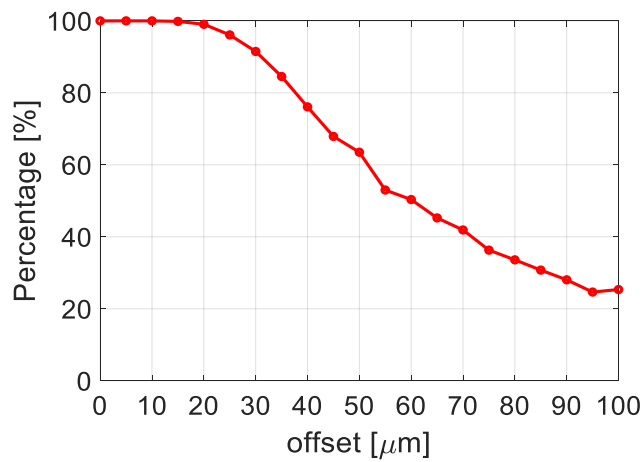

Figure 16: Orbit survival ratio due to FFT doublet offset

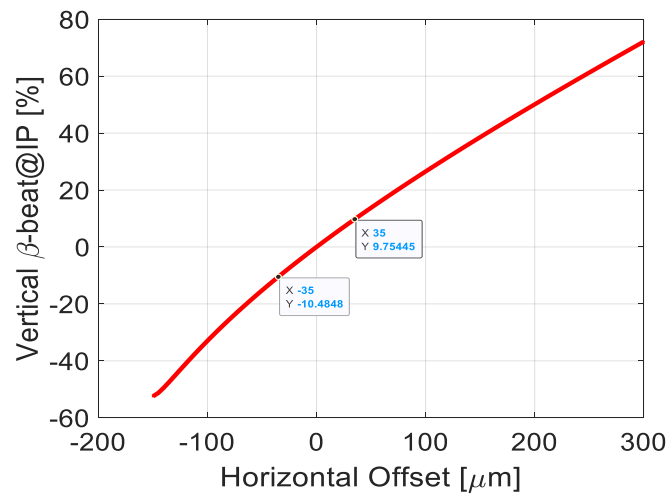

Figure 17: The vertical beta-beating induced by misalignments of the crab sextupoles

Table 3: The misalignment errors for the STCF collider rings

| Parameter | $\Delta x$ (μm) | $\Delta y$ (μm) | $\Delta s$ (μm) | $\Delta\theta$ (mrad) |
|---|---|---|---|---|
| Dipole | 100 | 100 | 100 | 0.1 |
| Quadrupole | 50 | 50 | 100 | 0.1 |
| FFT doublet | 30 | 30 | 100 | 0.1 |
| Arc/IR sextupoles | 50/30 | 50/30 | 100 | 0.1 |

Figure 18 illustrates the dynamic aperture of the collider rings with and without misalignment errors and corrections, both in the presence and absence of fringe field effects. Figure 19 presents the local momentum acceptance under the same conditions. The results show that the applied correction methods effectively restore the dynamic aperture to near-optimal levels, regardless of fringe field influences. However, the average momentum acceptance is slightly reduced with misalignment errors, decreasing from (+0.7%, −0.6%) to (+0.5%, −0.4%), which correspondingly reduces the Touschek lifetime from approximately 300 s to 230 s. Ongoing studies are focused on further improving the local momentum acceptance in the presence of misalignment errors to enhance beam lifetime and overall collider performance.

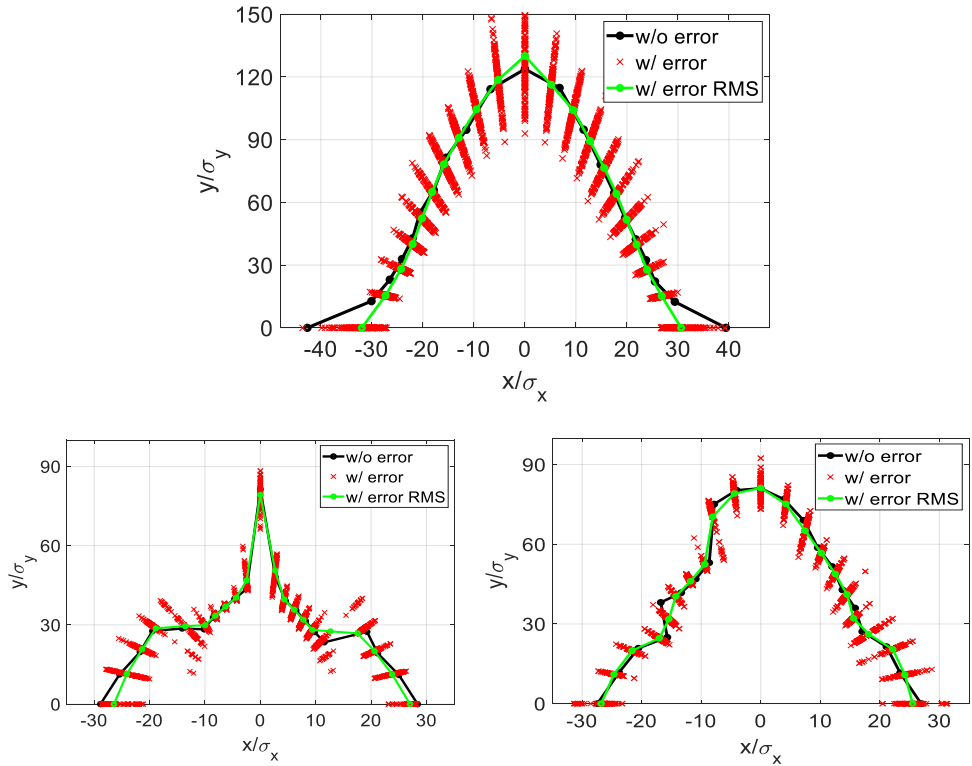

Figure 18: Dynamic aperture with misalignments and corrections, w/o and w/ fringe fields effect (Upper: DA without fringe fields; Lower left: DA with fringe fields effect and without corrections; Lower right: DA with fringe fields effect and octupole fields for correction).

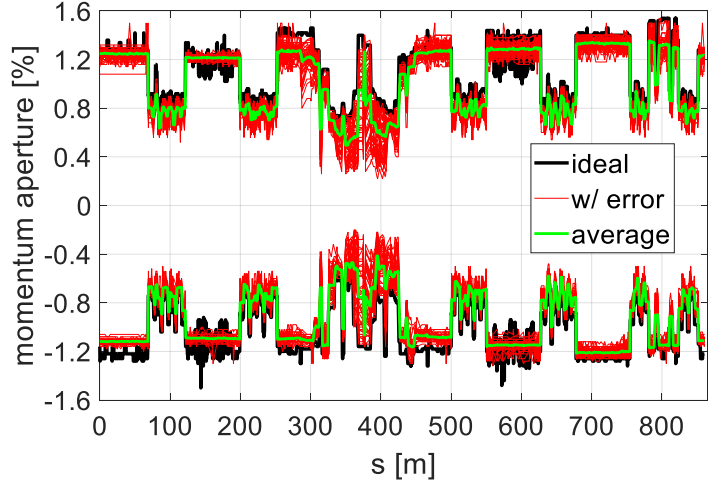

Figure 19: Momentum acceptance with and without misalignment errors and corrections with fringe fields effect and octupole fields for correction

## 8. CONCLUSIONS

The optics design of the STCF collider rings have been thoroughly investigated. A quasi-two-fold symmetric lattice has been proposed to optimize beam stability and performance. The modular IR incorporates local chromaticity correction up to the third order, following the design strategy pioneered by P. Raimondi. The Montague function is minimized at the IP and crab sextupole locations to enhance off-momentum beam dynamics. Additionally, the H-invariant function is minimized to maximize momentum acceptance, thereby improving the Touschek lifetime.

Nonlinear optimization is performed using the in-house developed PAMKIT code, complemented by frequency map analysis. Octupole compensation is applied to mitigate the effects of fringe fields in the final focus quadrupoles, ensuring robust lattice performance under misalignment and other errors. These techniques collectively enhance the collider's resilience to imperfections.

As a result for beam lifetime of longer than 300 s with ideal lattice and 230 s with errors and correction, we obtain a simulated luminosity of $9.4\times10^{34}$ $cm^{-2}s^{-1}$ at 4 GeV (CoM). Future design efforts will focus on refining nonlinear dynamics, optimizing collective effects, improving beam collimation, and addressing related challenges to achieve high luminosity and operational reliability.

## 9. ACKNOWLEDGMENTS

The authors would like to thank Anton Bogomyagkov and Mikhail Skamorokha from BINP, Kuanjun Fan and Cheng-Ying Tsai from HUST for very useful discussions and comments. This project is supported by the National Key R&D Program of China (Project No. 2022YFA1602201) and the National Natural Science Foundation of China (Project No. 12341501, No. 12405174). We also thank the Hefei Comprehensive National Science Center for the strong support on the STCF key technology research project.

**References**
[1] H. Peng, Y. Zheng and X. Zhou, Physics 49, 513 (2020).
[2] P. Raimondi, Status on Super-B Effort, Conf. Proc. C 0606141 (2006) 104.
[3] M. Zobov et al., Test of crab-waist collisions at DAFNE Phi factory, Phys. Rev. Lett. 104, 174801 (2010)

[4] A. Bogomyagkov, E. Levichev and P. Piminov, Final focus designs for crab waist colliders, Phys. Rev. Accel. Beams 19, 121005 (2016)

[5] Y. Ohnishi et al., Accelerator design at SuperKEKB, Prog. Theor. Exp. Phys. 2013, 03A011 (2013).

[6] Y. Ohnishi et al., SuperKEKB operation using crab waist collision scheme, Eur. Phys. J. Plus 136, 1023 (2021). https://doi.org/10.1140/epjp/s13360-021-01979-8

[7] SuperKEKB/Belle II Complete 2024 Operations, 10 January, 2025, https://www2.kek.jp/ipns/en/news/7015/

[8] CEPC Technical Design Report: Accelerator, Radiation Detection Technology and Methods, 8, 1-1105 (2024). https://doi.org/10.1007/s41605-024-00463-y

[9] FCC-ee: The Lepton Collider, Eur. Phys. J. Spec. Top. 228, 261–623 (2019). https://doi.org/10.1140/epjst/e2019-900045-4

[10] K. Oide et al., "Design of beam optics for the future circular collider e+e- collider rings", Phys. Rev. Accel. Beams 19, 111005 (2016)

[11] P. Raimondi, Local chromatic correction optics for Future Circular Collider e+e-, Phys. Rev. Accel. Beams 28, 021002 (2025)

[12] Tau/Charm Factory Accelerator Report, INFN-13-13/CLAB (2013)

[13] A. Bogomyagkov et al., Touschek lifetime and luminosity optimization for Russian Super Charm Tau factory, Journal of Instrumentation, 19 P02018 (2024)

[14] W.W. Gao et al., Interaction section lattice design for a STCF project, in Proceedings of the 10th International Particle Accelerator Conference, IPAC-2019, Melbourne, Australia (2019) doi:10.18429/JACoW-IPAC2019-MOPRB032

[15] J.Q. Lan et al., Design of beam optics for a Super Tau-Charm Factory, Journal of Instrumentation, 16 T07001 (2021)

[16] T. Liu et al., Recent progress and future plan for STCF collider ring lattice design, Modern Physics Letters A, Vol. 39, 2440005 (2024)

[17] L. Zhang et al., Longitudinal beam dynamics and collective effects at the STCF collider rings, Modern Physics Letters A, Vol. 39, 24440008 (2024)

[18] S. Li et al., Investigating luminosity optimization in STCF with crab waist scheme by beam-beam simulation, Modern Physics Letters A, Vol. 39, 2440013 (2024)

[19] N.S. Dikansky and D.V. Pestrikov, Effect of the crab waist and of the micro-beta on the beam-beam instability, Nucl. Instrum. Meth. A 600 (2009) 538

[20] A. Bogomyagkov, E. Levichev, and D. Shatilov, Beam-beam effects investigation and parameters optimization for a circular e+e− collider at very high energies, Phys. Rev. ST Accel. Beams 17, 041004 (2014)

[21] K. Ohmi et al., Coherent beam-beam instability in collisions with a large crossing angle, Physical Review Letters 119, 134801 (2017)

[22] K.L. Brown, A SECOND-ORDER MAGNETIC OPTICAL ACHROMAT, IEEE Transactions on Nuclear Science, Vol. NS-26, No. 3, June 1979

[23] K.L. Brown and R. Servranckx, First- and second-order charged particle optics, AIP Conf. Proc. 127 (1985) 62.

[24] MADX: "Methodical Accelerator Design", http://madx.web.cern.ch/madx.

[25] SAD Home Page, https://acc-physics.kek.jp/SAD/

[26] R. Servranckx and Karl L. Brown, Chromatic corrections for large storage rings, IEEE Trans.

Nucl. Sci. 26, 3598 (1979)

[27] B. Montague, Linear optics for improved chromaticity correction (1979), LEP Notes-1979, http://cds.cern.ch/record/67243.

[28] T. Liu, PAMKIT, https://pypi.org/project/PAMKIT/

[29] E. Levichev, P. Piminov, Analytic Estimation of the Non-Linear Tune Shift due to the Quadrupole Magnet Fringe Field, arXiv:0903.3028

[30] D. Robin et al., Global dynamics of the advanced light source revealed through experimental frequency map analysis, Phys. Rev. Lett. 85, 558 (2000)

[31] A. Bogomyagkov et al., Momentum acceptance optimization in FCC-ee lattice, in Proceedings of IPAC'16, Busan, Korea, 2016 (JACoW, Geneva, 2016), THPOR019.

[32] M. Borland, elegant: A Flexible SDDS-Compliant Code for Accelerator Simulation, Tech. Rep. LS-287, Advanced Photon Source (2000).